\documentclass[english,prl,twocolumn,groupedaddress,reprint]{revtex4}
\usepackage[T1]{fontenc}
\usepackage[latin9]{inputenc}
\usepackage{color}
\usepackage{amsmath}
\usepackage{amssymb}
\usepackage{graphicx}
\usepackage{esint}
\usepackage{comment}

\makeatletter
\@ifundefined{textcolor}{}
{%
 \definecolor{BLACK}{gray}{0}
 \definecolor{WHITE}{gray}{1}
 \definecolor{RED}{rgb}{1,0,0}
 \definecolor{GREEN}{rgb}{0,1,0}
 \definecolor{BLUE}{rgb}{0,0,1}
 \definecolor{CYAN}{cmyk}{1,0,0,0}
 \definecolor{MAGENTA}{cmyk}{0,1,0,0}
 \definecolor{YELLOW}{cmyk}{0,0,1,0}
}

\usepackage{bbold}
\usepackage{tikz}

\newcommand{\Ket}[1]{\left| #1 \right\rangle}

\newcommand{\reffig}[1]{Fig.~\ref{#1}}
\makeatother

\usepackage{babel}

\usepackage{minitoc}

\begin{document}

\title{Steady-State Entanglement in the Nuclear Spin Dynamics of a Double
Quantum Dot}

\author{M. J. A. Schuetz,$^{1}$ E. M. Kessler,$^{2,3}$ L. M. K. Vandersypen,$^{4}$ J. I. Cirac,$^{1}$ and G. Giedke$^{1}$ }

\affiliation{$^{1}$Max-Planck-Institut f\"ur Quantenoptik, Hans-Kopfermann-Str.
1, 85748 Garching, Germany}

\affiliation{$^{2}$Physics Department, Harvard University, Cambridge, MA 02318, USA}

\affiliation{$^{3}$ITAMP, Harvard-Smithsonian Center for Astrophysics, Cambridge, MA 02138, USA}

\affiliation{$^{4}$Kavli Institute of NanoScience, TU Delft, P.O. Box 5046, 2600 GA, Delft, The Netherlands}

\date{\today}
\begin{abstract}
We propose a scheme for the deterministic generation of steady-state
entanglement between the two nuclear spin ensembles in an electrically 
defined double quantum dot. 
Due to quantum interference in the collective coupling to the 
electronic degrees of freedom, the nuclear system is actively 
driven into a two-mode squeezed-like target state. 
The entanglement build-up is accompanied by a self-polarization
of the nuclear spins towards large Overhauser field gradients. 
Moreover, the feedback between the electronic and nuclear dynamics
leads to multi-stability and criticality in the steady-state solutions. 
\end{abstract}
\maketitle
Entanglement is a key ingredient
to applications in quantum information science.
In practice, however, it is very fragile and is often destroyed by the undesired
coupling of the system to its environment, hence robust ways
to prepare entangled states are called for. 
Schemes that exploit open system dynamics to prepare them 
as steady states are particularly promising  \cite{kraus04,verstraete09,diehl08,muschik11,sanchez13}. 
Here, we investigate such a scheme 
in quantum information architectures using spin qubits in quantum dots \cite{awschalom02,hanson07}.
In these systems, a great deal of research has been
directed towards the complex interplay between electron and nuclear
spins \cite{rudner11a,rudner11b,schuetz12,ono04,vink09,gullans10,chekhovich13,danon09},
with the ultimate goal of turning the nuclear spins from the dominant 
source of decoherence \cite{johnson05,koppens05,khaetskii02,bluhm10} into
a useful resource \cite{foletti09,taylor03,witzel07,ribeiro10}. 
The creation of entanglement between nuclear spins constitutes 
a pivotal element towards these goals. 

In this work, we propose a scheme for the dissipative
preparation of steady-state entanglement between the two nuclear spin
ensembles in a double quantum dot (DQD) 
in the Pauli-blockade regime \cite{ono02,hanson07}.
The entanglement arises from an interference between different
hyperfine-induced processes lifting the Pauli-blockade. 
This becomes possible by suitably engineering the effective
electronic environment, which ensures 
a \emph{collective} coupling of electrons
and nuclei (i.e., each flip can happen either in the left
or the right QD and no which-way information is leaked),
and 
that just two such processes with a common entangled stationary state are dominant.
Engineering of the electronic system via external gate
voltages 
facilitates the control of the
desired steady-state properties. 
Exploiting the
separation of electronic and nuclear
time-scales allows to derive a quantum master equation in which the
interference effect becomes apparent: It features non-local 
jump operators which drive the nuclear system into an entangled steady
state of EPR-type \cite{muschik11}.  Since the entanglement is actively stabilized by the
dissipative dynamics, our approach is inherently robust against weak random
perturbations \cite{kraus04,verstraete09,diehl08,muschik11,sanchez13}.  The
entanglement build-up is accompanied by a self-polarization of the nuclear
system towards large Overhauser (OH) field gradients 
if a small initial gradient is provided.
Upon surpassing a certain threshold value of this field 
the nuclear dynamics turn self-polarizing, and drive the
system to even larger gradients.  Entanglement is then
generated in the quantum fluctuations around these macroscopic
nuclear polarizations.  Furthermore, feedback between 
electronic and nuclear dynamics leads to multi-stability and criticality in
the steady-state solutions.

\begin{figure}
\includegraphics[width=1\columnwidth]{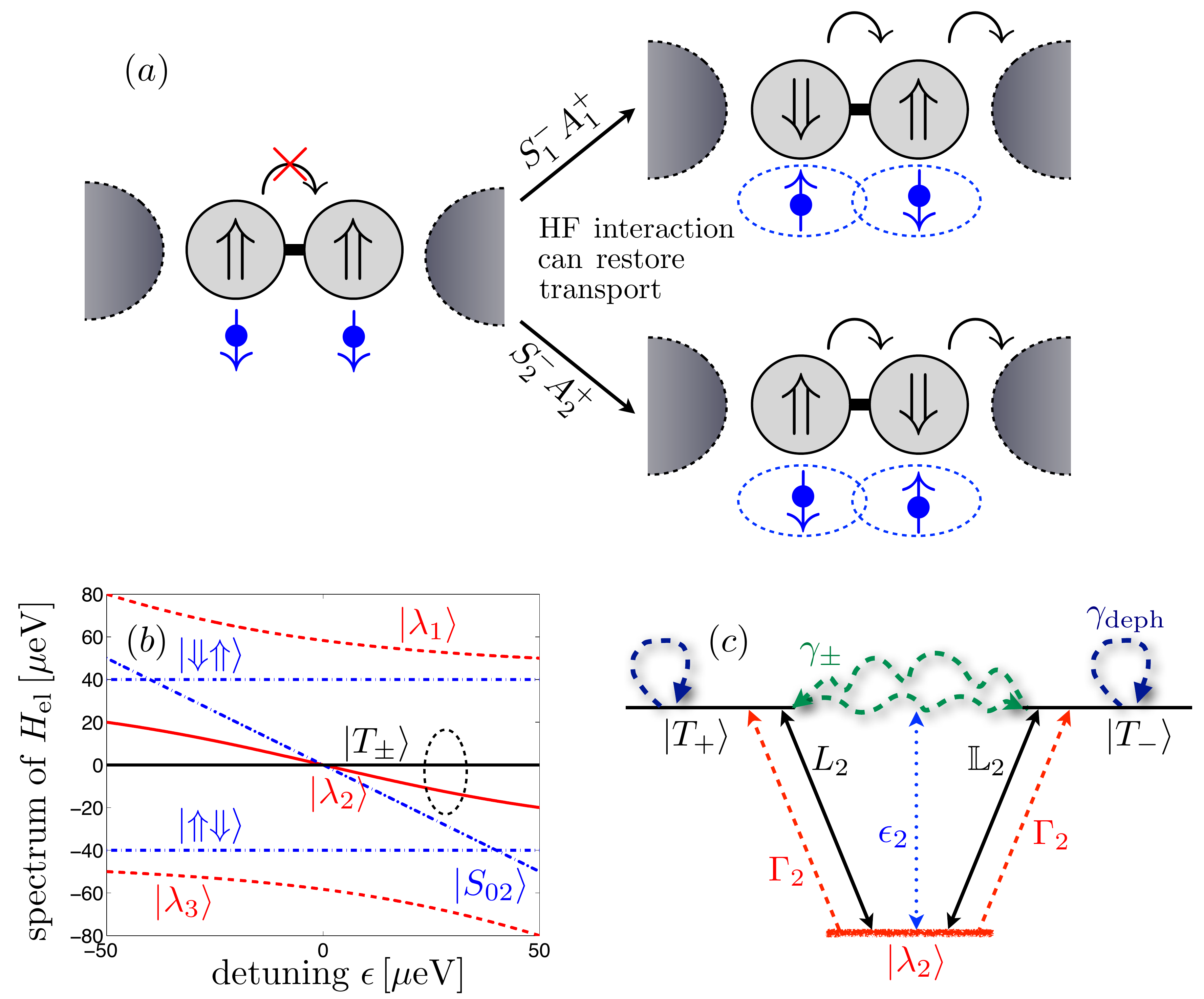}

\caption{\label{fig:Electronic-level-configuration}(color online). (a) Schematic
illustration of nuclear entanglement generation via electron transport.
Whenever the Pauli-blockade is lifted via the HF interaction with
the nuclear spins, a nuclear flip can occur in
either of the two dots. 
The local nature of the HF interaction
is masked by the non-local character of the electronic level $\left|\lambda_{2}\right\rangle $.
(b) Spectrum of $H_{\mathrm{el}}$ for $\Delta=40\mu\mathrm{eV}$ and $t=30\mu\mathrm{eV}$.
The three 
eigenstates $\left|\lambda_{k}\right\rangle $ are
displayed in red. The triplets $\left|T_{\pm}\right\rangle $ are
degenerate for $\omega_{0}=0$. In this setting, lifting of the spin
blockade due to HF interaction is pre-dominantly mediated by the non-local
jump operators required for two-mode squeezing, namely $L_{2}$ and
$\mathbb{L}_{2}$. The ellipse refers to a potential
operational area of our scheme. (c) The resulting effective
three-level system $\left\{ \left|T_{\pm}\right\rangle ,\left|\lambda_{2}\right\rangle \right\} $
including coherent HF coupling and the relevant dissipative
processes: $\left|\lambda_{2}\right\rangle $ decays according
to its overlap with 
$\left|S_{02}\right\rangle $
with an effective decay rate $\Gamma_{2}=\left|\left<\lambda_{2}|S_{02}\right>\right|^{2}\Gamma$ \cite{Gamma}.
Within this three-level subspace, purely electronic Pauli-blockade
lifting mechanisms like cotunneling 
or spin-orbital effects result
in effective dephasing and dissipative mixing rates, labeled as
$\gamma_{\mathrm{deph}}$ and $\gamma_{\pm}$, respectively.}
\end{figure}

We consider a DQD in the Pauli-blockade
regime \cite{ono02,hanson07}; see  \reffig{fig:Electronic-level-configuration}. 
A source-drain bias across the device
induces electron transport via the cycle $\left(0,1\right)\rightarrow\left(1,1\right)\rightarrow\left(0,2\right)\rightarrow\left(0,1\right)$.
Here, $\left(m,n\right)$ refers to a configuration with $m\left(n\right)$
electrons in the left (right) dot, respectively. 
The only energetically accessible $(0,2)$ state is the localized singlet, 
 $\Ket{S_{02}}$. Then, by the Pauli principle, 
the interdot charge transition $\left(1,1\right)\rightarrow\left(0,2\right)$ is 
allowed only for the $(1,1)$ spin-singlet 
$\left|S_{11}\right\rangle =\left(\left|\Uparrow\Downarrow\right\rangle -\left|\Downarrow\Uparrow\right\rangle \right)/\sqrt{2}$, 
while the spin-triplet states 
$\left|T_{\pm}\right\rangle $ and  $\left|T_{0}\right\rangle =\left(\left|\Uparrow\Downarrow\right\rangle +\left|\Downarrow\Uparrow\right\rangle \right)/\sqrt{2}$ 
are blocked.
Including a homogeneous Zeeman splitting $\omega_{0}$ 
and a magnetic gradient $\Delta$, both oriented along $\hat{z}$, the DQD within
the relevant two-electron subspace is then described by the
effective Hamiltonian $\left(\hbar=1\right)$ 
\begin{eqnarray}
H_{\mathrm{el}} & = & \omega_{0}\left(S_{1}^{z}+S_{2}^{z}\right)+\Delta\left(S_{2}^{z}-S_{1}^{z}\right)-\epsilon\left|S_{02}\right\rangle \left\langle S_{02}\right|\nonumber \\
 &  & +t\left(\left|\Uparrow\Downarrow\right\rangle \left\langle S_{02}\right|-\left|\Downarrow\Uparrow\right\rangle \left\langle S_{02}\right|+\mathrm{h.c.}\right),\label{eq:Two-electron-subspace-Hamiltonian}
\end{eqnarray}
where $\epsilon$ refers to the relative interdot energy detuning
between the left and right dot and $t$ describes interdot electron
tunneling in the Pauli-blockade regime.

The spin blockade inherent to $H_{\mathrm{el}}$ can be lifted, e.g.,
by the hyperfine (HF) interaction with nuclear spins in the host environment.
The electronic spins $\vec{S}_{i}$ confined in either of the two
dots $\left(i=1,2\right)$ are coupled to two different
sets of nuclei $\left\{ \sigma_{i,j}^{\alpha}\right\} $ via the isotropic
Fermi contact interaction \cite{schliemann03} 
\begin{equation}
H_{\mathrm{HF}}=\frac{a_{\mathrm{hf}}}{2}\sum_{i=1,2}\left(S_{i}^{+}A_{i}^{-}+S_{i}^{-}A_{i}^{+}\right)+a_{\mathrm{hf}}\sum_{i=1,2}S_{i}^{z}A_{i}^{z}.\label{eq:Hyperfine-Hamiltonian}
\end{equation}
Here, $S_{i}^{\alpha}$ and $A_{i}^{\alpha}=\sum_{j}a_{i,j}\sigma_{i,j}^{\alpha}$
for $\alpha=\pm,z$ denote electron and collective nuclear spin operators, and $a_{i,j}$ defines the unitless
HF coupling constant between the electron spin in dot $i$ and the
$j$th nucleus: $\sum_{j=1}^{N_{i}}a_{i,j}=N$, where 
$N=\left(N_{1}+N_{2}\right)/2\sim10^{6}$
refers to the average number of nuclei per dot. 
The individual nuclear spin operators 
$\sigma_{i,j}^{\alpha}$ 
are assumed to be spin-$\frac{1}{2}$
and we neglect the nuclear Zeeman and dipole-dipole
terms \cite{schliemann03}.
The second term in Eq.(\ref{eq:Hyperfine-Hamiltonian}) can be split 
into an effective nuclear magnetic field
and residual quantum fluctuations, $H_{\mathrm{zz}}=a_{\mathrm{hf}}\sum_{i=1,2}S_{i}^{z}\delta A_{i}^{z}$,
where $\delta A_{i}^{z}=A_{i}^{z}-\left\langle A_{i}^{z}\right\rangle _{t}$.
The (time-dependent) semiclassical OH field exhibits
a homogeneous 
$\bar{\omega}_{\text{\ensuremath{\mathrm{OH}}}}=\frac{a_{\mathrm{hf}}}{2}\left(\left\langle A_{1}^{z}\right\rangle _{t}+\left\langle A_{2}^{z}\right\rangle _{t}\right)$
and inhomogeneous component 
$\Delta_{\text{\ensuremath{\mathrm{OH}}}}=\frac{a_{\mathrm{hf}}}{2}\left(\left\langle A_{2}^{z}\right\rangle _{t}-\left\langle A_{1}^{z}\right\rangle _{t}\right)$,
which can be absorbed into the definitions of $\omega_{0}$ and $\Delta$
in Eq.(\ref{eq:Two-electron-subspace-Hamiltonian}) as $\omega_{0}=\bar{\omega}_{\text{\ensuremath{\mathrm{OH}}}}+\omega_{\mathrm{ext}}$
and $\Delta=\Delta_{\text{\ensuremath{\mathrm{OH}}}}+\Delta_{\mathrm{ext}}$,
respectively. For now, we assume the symmetric 
situation of vanishing external fields 
$\omega_{\mathrm{ext}}=\Delta_{\mathrm{ext}}=0$ \cite{SuppInfo}. 
Thus,  $\omega_{0}$ and $\Delta$ are dynamic variables depending on the nuclear polarizations.

The flip-flop dynamics, given by 
the first term in Eq.~(\ref{eq:Hyperfine-Hamiltonian})
and the OH fluctuations described by $H_{\mathrm{zz}}$
can be treated perturbatively with respect to the effective electronic
Hamiltonian $H_{\mathrm{el}}$. 
Its eigenstates within the $S_{\mathrm{tot}}^{z}=S_{1}^{z}+S_{2}^{z}=0$
subspace can be expressed as 
$\left|\lambda_{k}\right\rangle =\mu_k\left|\Uparrow\Downarrow\right\rangle +\nu_k\left|\Downarrow\Uparrow\right\rangle +\kappa_k\left|S_{02}\right\rangle $ ($k=1,2,3$) 
with corresponding eigenenergies $\epsilon_k$.
For $t \gg \omega_0, g_{\mathrm{hf}}$, where $g_{\mathrm{hf}}=\sqrt{N} a_\text{hf}$, $\Ket{\lambda_{1,3}}$ are far detuned, 
and the electronic subsystem can be simplified to an effective
three-level system comprising the levels $\left\{ \left|T_{\pm}\right\rangle ,\left|\lambda_{2}\right\rangle \right\} $.
Effects arising due to the presence of $\left|\lambda_{1,3}\right\rangle $
will be discussed below.
Within this reduced scheme, $H_{\mathrm{ff}}$ reads
\begin{equation}
H_{\mathrm{ff}}=\frac{a_{\mathrm{hf}}}{2}\left[L_{2}\left|\lambda_{2}\right\rangle \left\langle T_{+}\right|+\mathbb{L}_{2}\left|\lambda_{2}\right\rangle \left\langle T_{-}\right|+\mathrm{h.c.}\right],\label{eq:Hff-nonlocal-lambda2}
\end{equation}
where the \textit{non-local} nuclear operators $L_{2}=\nu_2A_{1}^{+}+\mu_2A_{2}^{+}$
and $\mathbb{L}_{2}=\mu_2A_{1}^{-}+\nu_2A_{2}^{-}$
are associated with lifting the Pauli-blockade from $\left|T_{+}\right\rangle $
and $\left|T_{-}\right\rangle $ via $\left|\lambda_{2}\right\rangle $,
respectively. 
They can be controlled via the external parameters $t$ and $\epsilon$ defining the amplitudes $\mu_2$ and $\nu_2$.

The dynamical evolution of the system is described
in terms of a Markovian master equation for the reduced
density matrix of the DQD system $\rho$ describing the relevant 
electronic and nuclear degrees of freedom \cite{schuetz12}.
Besides the HF
dynamics described above, it accounts for other purely electronic
mechanisms like, e.g., cotunneling. These effects and their implications 
for the nuclear dynamics are described in \cite{SuppInfo} 
and lead to effective decay and dephasing processes in the $T_\pm$ subspace
with rates $\gamma_\pm, \gamma_\mathrm{deph}$; see \reffig{fig:Electronic-level-configuration}~(c). 
For fast electronic dynamics ($\gamma_{\pm}, \gamma_{\mathrm{deph}} \gg g_{\mathrm{hf}}$) and
a sufficiently high gradient
$\Delta \gtrsim 3 \mu \mathrm{eV}$ (see \cite{SuppInfo}), the hybridized electronic level $\left|\lambda_{2}\right\rangle $
exhibits a significant overlap with the localized singlet $\left|S_{02}\right\rangle $ and
the  electronic subsystem settles in the desired quasi-steady state, 
$\rho_{\mathrm{ss}}^{\mathrm{el}}=\left(\left|T_{+}\right\rangle \left\langle T_{+}\right|+\left|T_{-}\right\rangle \left\langle T_{-}\right|\right)/2$,
on a time-scale much shorter than the nuclear dynamics. 
One can then adiabatically eliminate all electronic coordinates yielding 
a coarse-grained equation of motion for the nuclear density
matrix $\sigma=\mathsf{Tr}_{\mathrm{el}}\left[\rho\right]$, where
$\mathsf{Tr}_{\mathrm{el}}\left[\dots\right]$ denotes the trace over
the electronic degrees of freedom:
$\dot{\sigma}=\mathcal{L}_{\mathrm{id}}\left[\sigma\right]+\mathcal{L}_{\mathrm{nid}}\left[\sigma\right]$.
Here, the first dominant term describes 
the desired nuclear squeezing dynamics 
\begin{eqnarray}
\mathcal{L}_{\mathrm{id}}\left[\sigma\right] & = & \frac{\gamma}{2}\left[\mathcal{D}\left[L_{2}\right]\sigma+\mathcal{D}\left[\mathbb{L}_{2}\right]\sigma\right]\nonumber \\
 &  & +i\frac{\delta}{2}\left(\left[L_{2}^{\dagger}L_{2},\sigma\right]+\left[\mathbb{L}_{2}^{\dagger}\mathbb{L}_{2},\sigma\right]\right),\label{eq:QME-ideal-collective-spins}
\end{eqnarray}
where
$\mathcal{D}\left[c\right]\rho=c\rho c^{\dagger}-\frac{1}{2}\left\{ c^{\dagger}c,\rho\right\}$.
It arises from coupling to the level $\left|\lambda_{2}\right\rangle $,
while $\mathcal{L}_{\mathrm{nid}}\left[\sigma\right]$
results from coupling to the far detuned levels
$\left|\lambda_{1,3}\right\rangle $ and OH fluctuations described
by $H_{\mathrm{zz}}$  \cite{SuppInfo}. Here, $\gamma$ and $\delta$
refer to a HF-mediated decay rate and Stark shift, respectively
\footnote{
Microscopically, $\gamma$ and 
$\delta$ are given by 
$\gamma = a_{\mathrm{hf}}^{2} \tilde{\Gamma} / 2[\epsilon_{2}^2 + \tilde{\Gamma}^2]$ and
$\delta = (\epsilon_{2}/2\tilde{\Gamma})\gamma$, respectively. 
Here, $\tilde{\Gamma}=\Gamma_{2}+\gamma_{\pm}/2+\gamma_{\mathrm{deph}}/4$.
}.

\begin{figure}
\includegraphics[width=1\columnwidth]{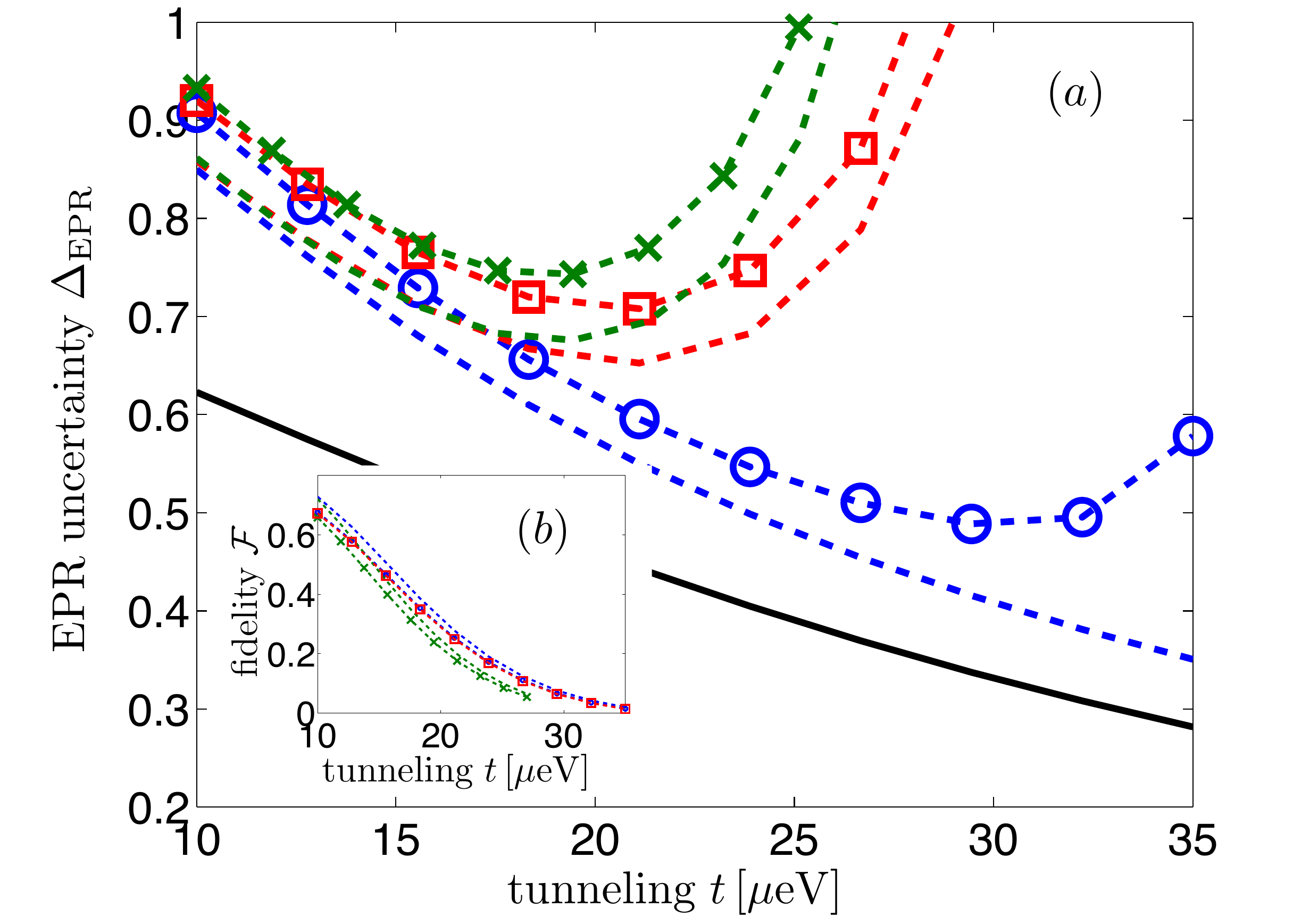}

\caption{\label{fig:steady-state-entanglement}(color online). Steady-state entanglement between
the two nuclear spin ensembles quantified via (a) the EPR-uncertainty
$\Delta_{\mathrm{EPR}}$ and (b) fidelity $\mathcal{F}$ 
of the nuclear steady state with the two-mode squeezed target state.
The black solid curve refers to the idealized setting
where the undesired HF coupling to $\left|\lambda_{1,3}\right>$ has been ignored and 
where $J_{1}=J_{2}=pJ_{\mathrm{max}}$, $p=0.8$ and $N_{1}=N_{2}=2J_{\mathrm{max}}=10^{6}$, 
corresponding to $\Delta_{\mathrm{OH}}=40\mu\mathrm{eV}$.
The blue-dashed line then also takes into account coupling to $\left|\lambda_{1,3}\right>$
while the red-dashed curve in addition accounts for an asymmetric dot size:
$N_{2}=0.8N_{1}=8\times10^{5}$. The amount of entanglement decreases
for a smaller nuclear polarization: $p=0.7$ (green dashed curve).
Classical uncertainty (symbols) in the total spin $J_{i}$ quantum 
numbers leads to less entanglement, but does not 
destroy it. Here, we have set the range of the distribution 
to $\Delta_{J_{i}}=50\sqrt{N_{i}}$.
Other numerical parameters: 
$\omega_{0}=0$, $\Gamma=25\mu\mathrm{eV}$, $\epsilon=30\mu\mathrm{eV}$ and
$\gamma_{\pm} + \gamma_{\mathrm{deph}}/2=1\text{\ensuremath{\mu}}\mathrm{eV}$.}
\end{figure}

Pure stationary solutions $\left| \xi_{\mathrm{ss}} \right\rangle$
associated with the dynamics generated by Eq.(\ref{eq:QME-ideal-collective-spins})
can be obtained from the dark-state condition 
$L_{2}\left| \xi_{\mathrm{ss}} \right\rangle=\mathbb{L}_{2}\left| \xi_{\mathrm{ss}} \right\rangle=0$.
First, we consider the limit of 
equal dot sizes $(N_{1}=N_{2})$ and
uniform HF coupling $\left(a_{i,j}=N/N_{i}\right)$,
and generalize our results later.
The nuclear system can be described via
Dicke states $\left|J_{i},k_{i}\right\rangle $,
where 
$k_{i}=0,1,\dots,2J_{i}$ and $J_{i}$ refer to the spin-$\hat{z}$ projection and total spin quantum numbers, respectively.
For $J_{1}=J_{2}=J$, one readily checks that the dark-state 
condition is satisfied by the (unnormalized) pure state 
$\left| \xi_{\mathrm{ss}} \right\rangle =\sum_{k=0}^{2J}\xi^{k}\left|J,k\right\rangle \otimes \left|J,2J-k\right\rangle $, 
representing an entangled state closely similar to the two-mode squeezed state \cite{SuppInfo}.
The parameter $\xi=-\nu_2/\mu_2$ quantifies the entanglement and polarization of the nuclear system.
$\left|\xi\right|<1$ ($\left|\xi\right|>1$) corresponds to states of large
positive (negative) OH gradients, respectively. 
The system is invariant under the symmetry 
transformation ($\mu_2 \leftrightarrow \nu_2$, $A_{1,2}^z \rightarrow - A_{1,2}^z$)
which gives rise to a bistability in the steady state, 
as for every solution with positive OH gradient ($\Delta>0$), 
we find another one with $\Delta<0$.

For a given $\left|\xi\right| \neq 1$ the individual 
nuclear polarizations in 
the state 
$\Ket{\xi_{\mathrm{ss}}}$  
approach one
as we increase the system size $J$, and we can describe the system dynamics in the vicinity of the respective steady state in the
framework of a Holstein-Primakoff (HP) transformation \cite{kessler12}.  
This allows for a detailed
analysis of the nuclear dynamics including perturbative effects 
from the processes described by $\mathcal L_\text{nid}$.
The collective nuclear spins $I_{i}^{\alpha}=\sum_{j}\sigma_{i,j}^{\alpha}$ are mapped to 
bosonic operators
\footnote{Here, we consider the subspace with large collective spin quantum numbers, 
$J_{i}\sim\mathcal{O}\left(N/2\right)$. The zeroth-order HP mapping can be justified
self-consistently, provided that the occupations in the bosonic modes 
are small compared to $2J_{i}$.}
and the (unique) ideal steady state is well-known to be a two-mode squeezed state \cite{muschik11, SuppInfo}
which represents 
$\left|\xi_{\mathrm{ss}}\right\rangle$ within the HP picture. 
Since in the bosonic case the modulus of $\xi$ is confined to $|\xi|<1$,
the HP analysis refers to one of the two symmetric steady-state solutions mentioned above.
Within the HP approximation the dynamics generated by 
$\dot{\sigma}=\mathcal{L}_{\mathrm{id}}\left[\sigma\right]+\mathcal{L}_{\mathrm{nid}}\left[\sigma\right]$
are quadratic in the new bosonic creation and annihilation
operators. Therefore, the nuclear dynamics are purely Gaussian
and exactly solvable. The generation of entanglement can be certified via 
the EPR entanglement condition \cite{muschik11,raymer03},
$\Delta_{\mathrm{EPR}}<1$, where 
$\Delta_{\mathrm{EPR}} = \left[\mathrm{var}\left(I_{1}^{x}+I_{2}^{x}\right)+\mathrm{var}\left(I_{1}^{y}+I_{2}^{y}\right)\right]/\left(\left|\left<I_{1}^{z}\right>\right|+\left|\left<I_{2}^{z}\right>\right|\right)$.  
While $\Delta_{\mathrm{EPR}} \geq 1$ for separable states, the ideal dynamics $\mathcal{L}_{\mathrm{id}}$
drive the nuclear spins into an EPR state with 
$\Delta_{\mathrm{EPR}}^{\mathrm{id}}=\left(1-\left|\xi\right|\right)/\left(1+\left|\xi\right|\right)<1$.
As illustrated in  \reffig{fig:steady-state-entanglement}, we numerically find
that the generation of steady-state entanglement persists even 
for asymmetric dot sizes of $\sim20\%$, 
classical uncertainty in the total spins $J_{i}$
\footnote{We average over an uniform distribution 
of $\{J_{1},J_{2}\}$ subspaces with a range of $\Delta_{J_{i}}=50\sqrt{N_{i}}$. 
The center of the distribution $\bar{J}_{i}$ has been taken as
$\bar{J}_{i}=pN_{i}/2$, where the polarization $p$ is set 
by the OH gradient via $p=\Delta_{\mathrm{OH}}/\Delta^{\mathrm{max}}_{\mathrm{OH}}$;
here, $\Delta^{\mathrm{max}}_{\mathrm{OH}}=A_{\mathrm{HF}}/2 \approx 50 \mu \mathrm{eV}$. }
and the undesired terms $\mathcal{L}_{\mathrm{nid}}$.
When tuning $t$ from $10\mu\mathrm{eV}$ 
to $35\mu\mathrm{eV}$, the squeezing parameter $|\xi|$ increases from 
$\sim0.2$ to $\sim0.6$, respectively.
For $|\xi|\approx0.2$, we obtain a relatively high fidelity $\mathcal{F}$ with 
the ideal two-mode squeezed state, close to 80$\%$.      
For stronger squeezing, the target state becomes more susceptible 
to the undesired noise terms, 
first leading to a reduction of $\mathcal{F}$ and
eventually to a break-down of the HP approximation.
The associated critical behavior 
can be understood in terms of a dissipative phase transition \cite{kessler12,schuetz13}.

We now turn to the experimental realization of our scheme \cite{SuppInfo}: 
In the analysis above, we discussed the idealized case of uniform HF coupling. 
However, our scheme also works 
for non-uniform coupling, provided that the two dots are sufficiently similar:
If the coupling is completely inhomogeneous, that is $a_{i,j} \neq a_{i,k}$ for all $j \neq k$, but the two QDs
are identical $\left(a_{1,j}=a_{2,j}\forall j=1,\dots,N_1\equiv N_2\right)$, 
Eq.(\ref{eq:QME-ideal-collective-spins}) supports a \textit{unique pure entangled} stationary state. 
Up to normalization, it reads
$\left|\xi_{\mathrm{ss}} \right\rangle =\otimes_{j=1}^{N}\left|\xi\right\rangle_{j} $, 
where $\left|\xi \right\rangle_{j}=\left|\downarrow_{j},\uparrow_{j}\right\rangle+\xi \left|\uparrow_{j},\downarrow_{j}\right\rangle$
is an entangled state of two nuclear spins belonging to different nuclear ensembles 
\footnote{Numerical evidence (for small systems) indicates that small deviations from perfect
symmetry between the QDs still yield
an entangled (mixed) steady state close to $\left|\xi_{\mathrm{ss}} \right\rangle$ \cite{SuppInfo}.}.
$\left|\xi_{\mathrm{ss}}\right>$ features a (large) polarization
gradient $\Delta_{I^{z}}=\left\langle I_{2}^{z}\right\rangle - \left\langle I_{1}^{z}\right\rangle = N\frac{1-\xi^2}{1+\xi^2}$.

\begin{figure}
\includegraphics[width=0.9\columnwidth]{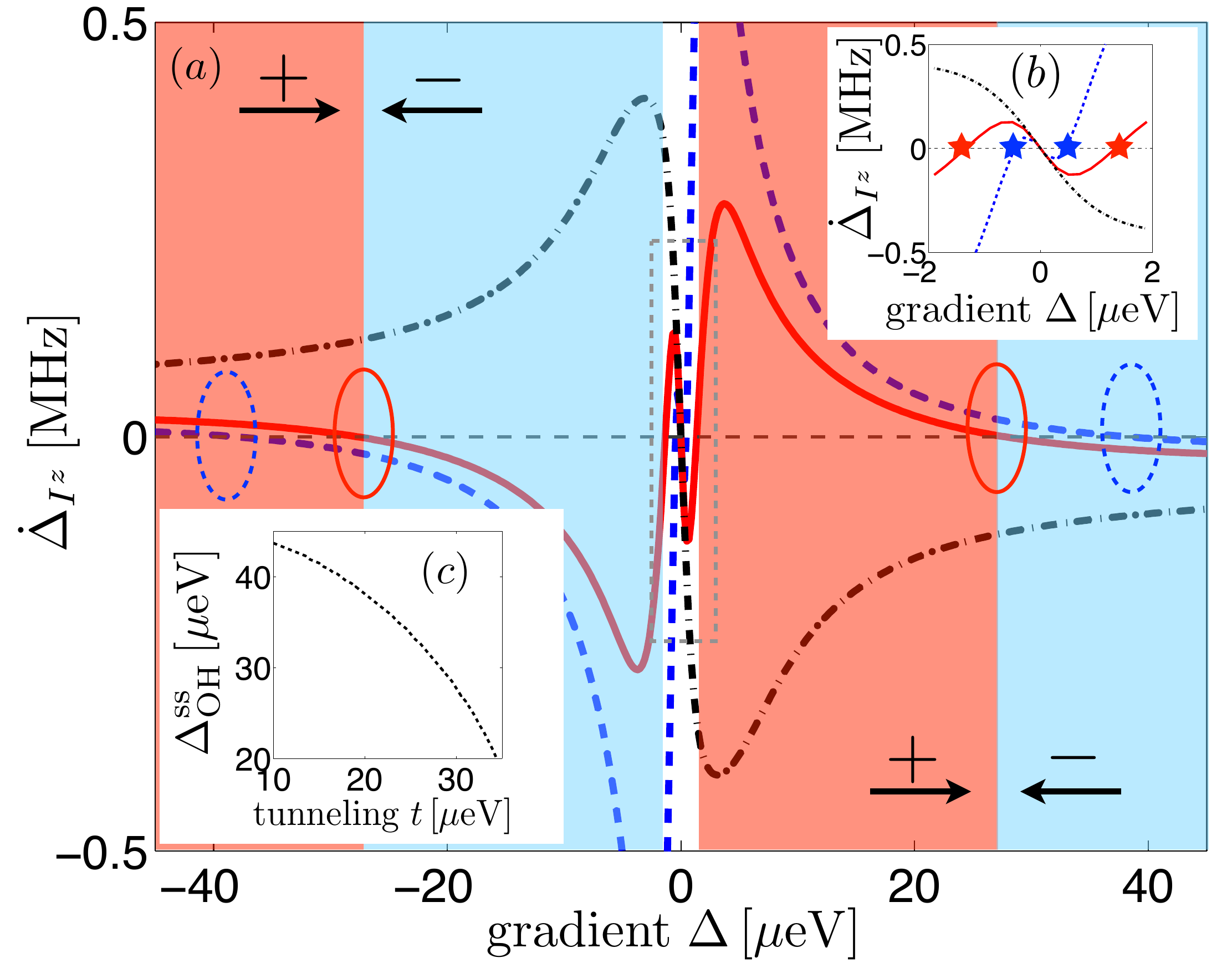}

\caption{\label{fig:Polarization-dynamics}(color online). Semiclassical solution
to the nuclear polarization dynamics. (a) Instantaneous nuclear polarization
rate $\dot{\Delta}_{I^{z}}$ as a function of the gradient $\Delta$
for $t=20\mu\mathrm{eV}$ (blue dashed), $t=30\mu\mathrm{eV}$ (red
solid) and $t=50\mu\mathrm{eV}$ (black dash-dotted). 
FPs are found at $\dot{\Delta}_{I^{z}}=0$. The ovals mark stable high-gradient
steady state solutions. The background coloring refers to the sign of $\dot{\Delta}_{I^{z}}$
(for $t=30\mu\mathrm{eV}$) which determines the stable 
FP the nuclear system is attracted
to (see arrows). (b) Zoom-in of (a) into the low-gradient regime:
The unpolarized FP 
lies at $\Delta=0$, whereas
critical, instable points $\Delta_{\mathrm{OH}}^{\mathrm{crt}}$
(marked by stars) can be identified with $\dot{\Delta}_{I^{z}}=0$
and $d\dot{\Delta}_{I^{z}}/d\Delta>0$. (c) Stable high-polarization FPs 
$\Delta_{\mathrm{OH}}^{\mathrm{ss}}$ (see ovals) as a function of
$t$; for $t\approx10\mu\mathrm{eV}$ we obtain a nuclear polarization
of $\sim90\%$. Other numerical parameters: $\Gamma=25\mu\mathrm{eV}$,
$\epsilon=30\mu\mathrm{eV}$, $\gamma_{\pm}=0.3\text{\ensuremath{\mu}}\mathrm{eV}$
and $\gamma_{\mathrm{deph}}=0.5\text{\ensuremath{\mu}}\mathrm{eV}$.
 }
\end{figure}

The build-up of a large OH gradient is corroborated 
within a semiclassical calculation which neglects correlations among
the nuclear spins \cite{SuppInfo}. 
This is valid on time scales long compared to nuclear
dephasing mechanisms  
\footnote{We estimate 
$\gamma_{\mathrm{eff}}^{-1}\approx 1\mathrm{s}$; 
this is compatible with the semiclassical approximation
and in agreement with typical polarization time scales \cite{gullans10, takahashi11}.} \cite{gullans10,christ07}. 
Assuming equal dot sizes, $N_{1}=N_{2}=N$, 
we use a semiclassical factorization scheme \cite{christ07} resulting in decoupled equations
of motion for the two nuclear polarization variables $\left\langle I_{1}^{z}\right\rangle _{t}$
and $\left\langle I_{2}^{z}\right\rangle _{t}$
\footnote{The results obtained within this approximative factorization scheme
have been confirmed by numerical simulations for small sets of nuclei \cite{schuetz13}.}. 
In particular, $\Delta_{I^{z}}$ evolves as 
\begin{eqnarray}
\frac{d}{dt}\Delta_{I^{z}} & = & -\gamma_{\mathrm{eff}}\left[\Delta_{I^{z}}-N\frac{\chi}{\gamma_{\mathrm{eff}}}\right],\label{eq:EOM-Delta-Iz-semiclassical-closed-1}
\end{eqnarray}
where the HF-mediated depolarization $\gamma_{\mathrm{eff}}$ and
pumping rate $\chi$
(see \cite{SuppInfo} for their connection to microscopic parameters)
depend on the gradient $\Delta$
defined in Eq.~(\ref{eq:Two-electron-subspace-Hamiltonian}),
in particular on the OH gradient $\Delta_{\mathrm{OH}}\propto\Delta_{I^{z}}$.
The electron-nuclear feedback-loop can then be closed self-consistently
by identifying steady-state solutions of Eq.~(\ref{eq:EOM-Delta-Iz-semiclassical-closed-1})
in which the parameter $\Delta$ is provided by the nuclear OH
gradient only. The instantaneous polarization rate $\dot{\Delta}_{I^{z}}$,
given in Eq.~(\ref{eq:EOM-Delta-Iz-semiclassical-closed-1}), is displayed
in \reffig{fig:Polarization-dynamics} as a function of $\Delta$,
with the electronic subsystem in its respective steady state, yielding
a non-linear equation for the nuclear equilibrium polarizations. 
Stable fixed points (FPs) are determined by $\dot{\Delta}_{I^{z}}=0$
and $d\dot{\Delta}_{I^{z}}/d\Delta<0$ as opposed to instable ones
where $d\dot{\Delta}_{I^{z}}/d\Delta>0$ \cite{vink09,danon09,bluhm10b}. 
We can identify parameter regimes in which the nuclear 
system features three FPs which are interspersed by two instable points. 
Two of the stable FPs are high-polarization solutions of opposite sign,
supporting a macroscopic OH gradient, while one is the trivial, zero
polarization solution. If the initial gradient lies outside the instable points, the 
system turns self-polarizing and the OH gradient 
approaches a highly polarized FP.   
For typical parameter values we estimate that the OH gradient 
at the instable points is $\approx(1-2)\mu\mathrm{eV}$; compare \reffig{fig:Polarization-dynamics} (b).
This comparatively moderate initial gradient could be achieved via, e.g., 
a nanomagnet \cite{pioro08, petersen13} or alternative 
dynamic nuclear polarization schemes \cite{foletti09, gullans10, petta08, takahashi11}.

Next, we address the effects of weak nuclear interactions:
First, we have neglected nuclear dipole-dipole interactions. 
However, we estimate the time scale for
the entanglement creation as 
$t^{*} = \hbar/N\gamma \lesssim10\mu\mathrm{s}$
which is fast compared to typical nuclear decoherence times, recently
measured to be $\sim1\mathrm{ms}$ in vertical DQDs 
\cite{takahashi11}. Thus, it should be possible to create entanglement
between the two nuclear spin ensembles faster than it gets disrupted
by dipole-dipole interactions among the nuclei.
Second, we have disregarded 
nuclear Zeeman terms since 
our scheme requires no external homogeneous magnetic field 
for sufficiently strong tunneling $t$ \footnote{Note that any 
initial OH splitting $\bar{\omega}_{\mathrm{OH}}$ is damped to zero in the
steady state \cite{schuetz13}.}. 

Finally, entanglement could be detected by measuring the OH shift in each dot
separately \cite{hanson07}; in combination with NMR techniques
to rotate the nuclear spins \cite{chekhovich13} we can
obtain all spin components and their variances which are
sufficient to verify the presence of entanglement (similar to the proposal \cite{rudner11a}).

To conclude, we have presented a scheme for the dissipative entanglement
generation among the two nuclear spin ensembles in a DQD. This may 
provide a long-lived, solid-state entanglement resource and a new route 
for nuclear-spin-based information storage and manipulation.

\textit{Acknowledgments.}---We
acknowledge support by 
the DFG within SFB 631, the Cluster of Excellence NIM and
the project MALICIA within the 7th Framework Programme for Research of
the European Commission, under FET-Open grant number: 265522.
EMK acknowledges support by the Harvard Quantum Optics Center
and the Institute for Theoretical Atomic and Molecular Physics.
LV acknowledges support by the Dutch Foundation for Fundamental Research on Matter (FOM).

\newpage

\appendix

\section{Supplementary Information (SI)}

The following supplementary information (SI) provides additional background 
material to specific topics of the main text. 
First, we discuss the master equation used to model the dynamics of the DQD. 
Then, by eliminating all electronic coherences, we derive an effective description
for the nuclear dynamics. 
Thereafter, it is shown that this description can be simplified substantially
in the high gradient regime where the electronic level $\left|\lambda_{2}\right>$
can be eliminated from the dynamics.
The explicit form of the noise terms labeled by $\mathcal{L}_{\mathrm{nid}}$
in the main text is given thereafter. 
The following section presents analytical and numerical results on the ideal nuclear 
target state, for both uniform and non-uniform HF coupling.  
Next, we present details on the Holstein-Primakoff mapping 
and give the so-called standard form of the covariance matrix
which has been used for the evaluation of the 
EPR uncertainty $\Delta_{\mathrm{EPR}}$ within the HP approximation.
Finally, we provide some details regarding our semiclassical approach to 
study the nuclear self-polarization effects, discuss the effect of external magnetic fields
and summarize the requirements for an experimental realization of our scheme.

\tableofcontents

\subsection{The Model \label{the-model}}

After tracing out the unobserved degrees of
freedom of the leads,
the dynamical evolution of the system can be described in terms of an
effective Markovian master equation for the reduced density matrix
of the DQD system $\rho$ describing the relevant electronic as well
as the nuclear subsystem. Within the relevant three-level subspace 
$\{\left|T_{\pm}\right\rangle, \left|\lambda_{2}\right\rangle\}$, it reads 
\begin{eqnarray}
\dot{\rho} & = & \mathcal{L}_{0}\left[\rho\right]+\mathcal{V}\left[\rho\right] \label{eq:QME-1-1}\\
 \mathcal{L}_{0}\left[\rho\right] & = & -i\left[H_{\mathrm{el}},\rho\right]+\Gamma_{2}\sum_{\nu=\pm}\mathcal{D}\left[\left|T_{\nu}\right\rangle \left\langle \lambda_{2}\right|\right]\rho \nonumber \\
 &  & +\gamma_{\pm}\sum_{\nu=\pm}\mathcal{D}\left[\left|T_{\bar{\nu}}\right\rangle \left\langle T_{\nu}\right|\right]\rho+\mathcal{L}_{\mathrm{deph}}\left[\rho\right],\nonumber 
\end{eqnarray}
where $\mathcal{V}\left[\rho\right]=-i\left[H_{\mathrm{ff}}+H_{\mathrm{zz}},\rho\right]$
and $\mathcal{D}\left[c\right]\rho$ is a short-hand notation for
the Lindblad term $\mathcal{D}\left[c\right]\rho=c\rho c^{\dagger}-\frac{1}{2}\left\{ c^{\dagger}c,\rho\right\} $.
In deriving Eq.(\ref{eq:QME-1-1}), we have neglected terms rotating at 
a frequency of $\epsilon_{l}-\epsilon_{k}$ for $k\neq l$ and dissipative 
terms acting entirely within the fast subspace, i.e., terms of the form
$\mathcal{D}\left[\left|\lambda_{k}\right\rangle \left\langle \lambda_{j}\right|\right]$;
for typical parameters, we have checked that the simplified Liouvillian given in Eq.(\ref{eq:QME-1-1})
reproduces exactly the electronic quasi steady state (fulfilling $\mathcal{L}_{0}\left[\rho_\mathrm{ss}^{\mathrm{el}}\right]=0$).
Moreover,  it describes very well the electronic asymptotic decay rate, that is 
the spectral gap of $\mathcal{L}_{0}$, which quantifies the long-time 
behavior of the electronic subsystem \cite{kessler12, schuetz13} and is therefore
relevant for a good description of the nuclear dynamics.

\textit{Electron transport}.---Apart from the unitary dynamics discussed in the main text, Eq.(\ref{eq:QME-1-1})
contains three dissipative terms: 
The first one, proportional to 
$\Gamma_{2}=\left|\left<\lambda_{2}|S_{02}\right>\right|^{2}\Gamma$,
describes electron transport as the hybridized level $\left|\lambda_{2}\right\rangle $
acquires a finite lifetime according to its overlap with the localized
singlet $\left|S_{02}\right\rangle $. 
Here, $\Gamma$ is given by 
\begin{eqnarray}
\Gamma=\Gamma_{R}/2,
\end{eqnarray}
where
\begin{eqnarray}
\Gamma_{\alpha}=2\pi |T_{\alpha}|^2 n_{\alpha},
\end{eqnarray}
denotes the typical sequential tunneling rate to the lead $\alpha=L,R$; the tunnel matrix element 
$T_{\alpha}$ specifies the transfer coupling between the lead $\alpha$ and the DQD system
and $n_{\alpha}$ refers to the density of states per spin in the lead $\alpha$ \cite{schuetz12}.
By making the left tunnel barrier more transparent than the right one $(2\Gamma_{L} \gg \Gamma_{R})$,
we can eliminate the intermediate stage in the sequential tunneling process 
$\left(0,2\right)\rightarrow\left(0,1\right)\rightarrow\left(1,1\right)$ \cite{petersen13, schuetz12}.    
Then, on relevant time scales, the DQD is always in the two-electron regime 
and electron transport is fully described by the effective rate $\Gamma$. 


\textit{Other mechanisms}.---The second and third
dissipative term account for decay processes from $\left|T_{+}\right\rangle $
to $\left|T_{-}\right\rangle $ and vice versa and dephasing between
the triplets $\left|T_{\pm}\right\rangle $ which is modeled by the
Lindblad term 
\begin{eqnarray}
\mathcal{L}_{\mathrm{deph}}\left[\rho\right]=\frac{\gamma_{\mathrm{deph}}}{2}\mathcal{D}\left[\left|T_{+}\right\rangle \left\langle T_{+}\right|-\left|T_{-}\right\rangle \left\langle T_{-}\right|\right]\rho.
\end{eqnarray}
For the sake of theoretical generality, this is a common phenomenological
description for distinct physical mechanisms like e.g. cotunneling,
spin-exchange with the leads or spin-orbital effects which may also
lift the Pauli blockade and therefore contribute to electron transport
through the DQD device, but, in contrast to the HF interaction, do
so without affecting the nuclear spins directly. In accordance with
a typical experimental situation, they are weak compared to direct
tunneling in the singlet subspace, but may still be fast compared
to the typical HF time scale $g_{\mathrm{hf}}=A_{\mathrm{HF}}/\sqrt{N}\approx0.1\mu\mathrm{eV}$.

Our regime of interest can be summarized as 
\begin{eqnarray}
1 \gg \gamma_{\pm}/\Gamma,\gamma_{\mathrm{deph}}/\Gamma \gg g_{\mathrm{hf}}/\Gamma. \label{eq:regime-of-interest}
\end{eqnarray}
The right hand side 
can be suppressed efficiently
by working in a regime of strong electron exchange with the leads.
For typical values, we estimate $g_{\mathrm{hf}}/\Gamma \approx (2-4) \times 10^{-3}$.
In particular, this condition allows us to adiabatically eliminate all electronic coherences
for $\gamma_{\pm}+\gamma_{\mathrm{deph}}/2 \gg g_{\mathrm{hf}}$ and, in the
high gradient regime specified below, all
electronic coordinates can be eliminated for $2\gamma_{\pm} \gg g_{\mathrm{hf}}$.

\textit{Cotunneling}.---For example, let us briefly show how virtual tunneling
processes via 
localized
triplet states fit into this effective, phenomenological description.  
Usually, they are neglected because they
are far off in energy due to the relatively large singlet-triplet
splitting $\Delta_{\mathrm{st}} \gtrsim 400\mu\mathrm{eV}$ \cite{hanson07}. 
Still, they may contribute to electron transport by lifting the spin
blockade as follows: The triplet $\left|T_{\pm}\right\rangle $ with
$\left(1,1\right)$ charge configuration is coherently coupled to the localized triplet
$\left|T_{\pm}\left(0,2\right)\right\rangle $ by the interdot tunneling
coupling $t$. This transition is strongly detuned by the singlet-triplet
splitting $\Delta_{\mathrm{st}}$. Once, the energetically high lying
level $\left|T_{\pm}\left(0,2\right)\right\rangle $ is populated, it
quickly decays with an effective rate $\Gamma$ either back to $\left|T_{\pm}\right\rangle $
giving rise to pure dephasing or to $\left|T_{\mp}\right\rangle $ via
some fast intermediate steps. The former contributes to $\gamma_{\mathrm{deph}}$,
while the latter can be absorbed into the phenomenological rate $\gamma_{\pm}$. 
Using standard second order perturbation theory,
the effective rate for this mechanism can be estimated as 
\begin{eqnarray}
\gamma_{\mathrm{ct}} / \Gamma =x_{\mathrm{ct}} \approx \left(\frac{t}{\Delta_{\mathrm{st}}}\right)^2. 
\end{eqnarray}
Compared to direct electronic processes, it is lowered by the 'penalty' factor $x_{\mathrm{ct}}$, 
which we estimate as $x_{\mathrm{ct}} \approx (30/400)^2 \approx 0.005$.

\textit{Spin-orbit}.---Other mechanisms besides cotunneling also contribute to the 
phenomenological rates $\gamma_{\mathrm{deph}}$ and $\gamma_{\pm}$. 
For example, along the lines of the cotunneling analysis, spin-orbital effects can be accounted for. 
The corresponding penalty factor can be estimated as 
\begin{eqnarray}
x_{\mathrm{so}} \approx \frac{t_{\mathrm{so}}^2}{\epsilon^2 + \Gamma^2},
\end{eqnarray}
where the spin-orbit coupling parameter is approximately 
$t_{\mathrm{so}} \approx (0.01-0.1)t$ \cite{schreiber11, giavaras13}. 
This gives the order-of-magnitude estimate $x_{\mathrm{so}} \approx 3^2 / (30^2 + 25^2) \approx 0.006$.


On a similar footing, one can also account for spin-exchange with the leads \cite{schuetz13}. 
The different electronic decay channels have to be summed up as
$\gamma_{\pm}=\gamma_{\mathrm{ct}}+\gamma_{\mathrm{so}}+\dots$ and
$\gamma_{\mathrm{deph}}=\gamma_{\mathrm{ct}}+\gamma_{\mathrm{so}}+\dots$. 
Based on the estimates stated above, sufficiently strong electron exchange with the leads 
ensures the validity of Eq.(\ref{eq:regime-of-interest}). 

\subsection{Effective Nuclear Dynamics}

In the limit $\gamma_{\text{\ensuremath{\pm}}}+\gamma_{\mathrm{deph}}/2 \gg g_{\mathrm{hf}}$,
any electronic coherences decay rapidly on typical nuclear time scales.
Using standard techniques, we can then adiabatically eliminate them
from the dynamics yielding a simplified coarse-grained equation of
motion for the nuclear density matrix $\sigma=\mathsf{Tr}_{\mathrm{el}}\left[\rho\right]$,
where $\mathsf{Tr}_{\mathrm{el}}\left[\dots\right]$ denotes the trace
over the electronic degrees of freedom. 
Since differences in the populations of the triplets $\left|T_{+}\right>$
and $\left|T_{-}\right>$ are quickly damped to zero with a rate of $2\gamma_{\pm}$, 
it is approximately given by 
\begin{eqnarray}
\dot{\sigma} & = & \gamma\left\{ p_{+}\left[\mathcal{D}\left[L_{2}\right]\sigma+\mathcal{D}\left[\mathbb{L}_{2}\right]\sigma\right]\right.\label{eq:EOM-nuclear-spins-1}\\
 &  & \left.+\left(1-2p_{+}\right)\left[\mathcal{D}\left[L_{2}^{\dagger}\right]\sigma+\mathcal{D}\left[\mathbb{L}_{2}^{\dagger}\right]\sigma\right]\right\} \nonumber \\
 &  & +i\delta \left\{ p_{+}\left(\left[L_{2}^{\dagger}L_{2},\sigma\right]+\left[\mathbb{L}_{2}^{\dagger}\mathbb{L}_{2},\sigma\right]\right)\right.\nonumber \\
 &  & \left.-\left(1-2p_{+}\right)\left(\left[L_{2}L_{2}^{\dagger},\sigma\right]+\left[\mathbb{L}_{2}\mathbb{L}_{2}^{\dagger},\sigma\right]\right)\right\} ,\nonumber 
\end{eqnarray}
where $\gamma$ and $\delta$ refer to the effective
rate 
\begin{eqnarray}
\gamma=\frac{a_{\mathrm{hf}}^{2}\tilde{\Gamma}}{2[\tilde{\Gamma}^{2}+\epsilon_{2}^{2}]}
\end{eqnarray}
and Stark shift 
\begin{eqnarray}
\delta=(\epsilon_{2}/2\tilde{\Gamma})\gamma,
\end{eqnarray}
respectively. Here, we have set  
$\tilde{\Gamma}=\Gamma_{2}+\gamma_{\pm}/2+\gamma_{\mathrm{deph}}/4$.
The nuclear dynamics are governed by the \textit{non-local}
jump operators $L_{2}$ and $\mathbb{L}_{2}$, describing HF-mediated
nuclear flips from $\left|T_{+}\right\rangle $ and $\left|T_{-}\right\rangle $
to $\left|\lambda_{2}\right\rangle $, respectively, but still coupled
to the electronic subsystem via the population of the triplet $\left|T_{+}\right\rangle $,
$p_{+}$. 
On a coarse-grained time scale relevant for the nuclear
dynamics, all electronic coherences are fully depleted and the populations
(given by $p_{+}$, $p_{-}=p_{+}$ and
$p_{2}=1-2p_{+}$, respectively) 
completely characterize the electronic subsystem. 
Therefore, the coupled electron-nuclear DQD system is described
by Eq.(\ref{eq:EOM-nuclear-spins-1}), complemented by an equation
of motion for $p_{+}$, which, in turn, depends on the state of the nuclear spins \cite{schuetz13}. 
In Eq.(\ref{eq:EOM-nuclear-spins-1}) we have suppressed contributions arising
from the OH fluctuations, governed by $H_{\mathrm{zz}}$. This is in line
with the semiclassical approximation to study the nuclear polarization dynamics.
However, as stated in the main text, they have been taken into account when 
analyzing the steady state entanglement properties of the nuclear system (see below).

\subsection{High Gradient Regime}

\begin{figure}
\includegraphics[width=1\columnwidth]{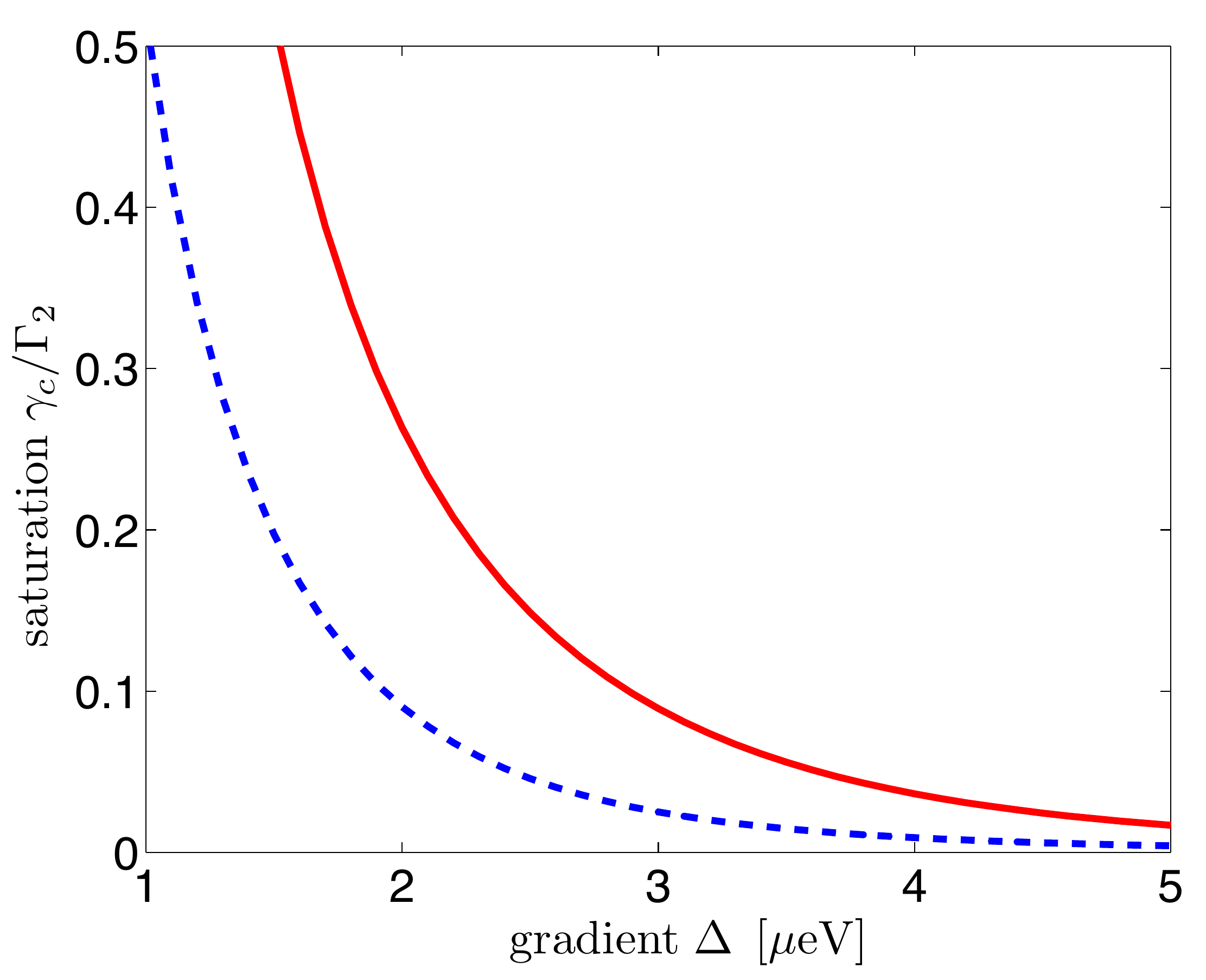}

\caption{\label{fig:saturation-depletion-lambda2}(color online). Saturation
parameter $\gamma_{c}/\Gamma_{2}$ as a function of the gradient $\Delta$
for $t=20\mu\mathrm{eV}$ (blue dashed) and $t=30\mu\mathrm{eV}$
(red solid), respectively. In the high gradient regime, where this
value is sufficiently low, the electronic level $\left|\lambda_{2}\right\rangle $
can be eliminated adiabatically from the dynamics as it gets fully
depleted on relevant nuclear time scales. Other numerical parameters
are: $\Gamma=25\mu\mathrm{eV}$, $\epsilon=30\mu\mathrm{eV}$, $\gamma_{\mathrm{deph}}=0.5\mu\mathrm{eV}$
and $\gamma_{\pm}=0.3\mu\mathrm{eV}$. }
\end{figure}

For a sufficiently high gradient
$\Delta$, the electronic level $\left|\lambda_{2}\right\rangle $
exhibits a significant overlap with the localized singlet $\left|S_{02}\right\rangle $;
accordingly, in this regime the electronic degrees of freedom 
can be eliminated completely from the
dynamics. More rigorously, this holds for 
\begin{equation}
\frac{\gamma_{c}}{\Gamma_{2}}=\frac{g_{\mathrm{hf}}^{2}}{2\left[\tilde{\Gamma}^{2}+\epsilon_{2}^{2}\right]}\frac{\tilde{\Gamma}}{\Gamma_{2}}\ll1,
\end{equation}
where $\gamma_{c}=N\gamma$ comprises a factor of $N\approx10^{6}$
to account for typical HF-mediated interaction strengths of $g_{\mathrm{hf}}=\sqrt{N} a_{\mathrm{hf}}\approx0.1\mu\mathrm{eV}$.
As shown in \reffig{fig:saturation-depletion-lambda2}, for typical
parameters $\left|\lambda_{2}\right\rangle $ can be eliminated adiabatically
for $\Delta\gtrsim\left(2-3\right)\mu\mathrm{eV}$. In this regime,
the electronic subsystem settles to a quasi steady-state, given by 
$\rho_{\mathrm{ss}}^{\mathrm{el}}=\left(\left|T_{+}\right\rangle \left\langle T_{+}\right|+\left|T_{-}\right\rangle \left\langle T_{-}\right|\right)/2$,
on a time scale much shorter than the nuclear dynamics. The effective 
nuclear dynamics in the submanifold of this electronic quasi 
steady state $\rho_{\mathrm{ss}}^{\mathrm{el}}$
gives rise to the Liouvillian
stated in the main text in Eq.(4).

\subsection{Noise Terms}
For completeness, here we present the explicit form of the superoperator 
$\mathcal{L}_{\mathrm{nid}}\left[\sigma\right]$ which can be decomposed as
\begin{equation}
\mathcal{L}_{\mathrm{nid}}\left[\sigma\right]=\mathcal{K}_{\mathrm{fz}}\left[\sigma\right]+\mathcal{K}^{\mathrm{nid}}_{\mathrm{ff}}\left[\sigma\right]+\mathcal{K}_{\mathrm{zz}}\left[\sigma\right].
\end{equation}
The first term is given by
\begin{eqnarray}
\mathcal{K}_{\mathrm{fz}}\left[\sigma\right] & = & -i\frac{a_{\mathrm{hf}}}{2}\sum_{i,\alpha=\pm}\left\langle S_{i}^{\alpha}\right\rangle _{\mathrm{ss}}\left[A_{i}^{\bar{\alpha}},\sigma\right] \nonumber \\
 &  & -ia_{\mathrm{hf}}\sum_{i}\left\langle S_{i}^{z}\right\rangle _{\mathrm{ss}}\left[\delta A_{i}^{z},\sigma\right].
\end{eqnarray}
Here, $\left\langle \cdot\right\rangle _{\mathrm{ss}}=\mathsf{Tr}_{\mathrm{el}}\left[\cdot\rho_{\mathrm{ss}}^{\mathrm{el}}\right]$
denotes the steady state expectation value. 
Next, undesired, second-order HF-mediated transitions to the 
electronic levels  $\left|\lambda_{1,3}\right>$ are described by 
\begin{eqnarray}
\mathcal{K}^{\mathrm{nid}}_{\mathrm{ff}}\left[\sigma\right] & = & \sum_{k\neq2}\left[\frac{\gamma_{k}}{2}\mathcal{D}\left[L_{k}\right]\sigma+i\frac{\delta_{k}}{2}\left[L_{k}^{\dagger}L_{k},\sigma\right]\right. \nonumber \\
 &  & \left.+\frac{\gamma_{k}}{2}\mathcal{D}\left[\mathbb{L}_{k}\right]\sigma+i\frac{\delta_{k}}{2}\left[\mathbb{L}_{k}^{\dagger}\mathbb{L}_{k},\sigma\right]\right],
\end{eqnarray}
where we have introduced the generalized, effective HF-mediated decay rates 
\begin{eqnarray}
\gamma_{k} & = & \frac{a_{\mathrm{hf}}^{2}\tilde{\Gamma}_{k}}{2\left[\epsilon_{k}^{2}+\tilde{\Gamma}_{k}^{2}\right]},
\end{eqnarray}
with the dephasing rate $\tilde{\Gamma}_{k}=\Gamma_{k}+\gamma_{\mathrm{\pm}}/2+\gamma_{\mathrm{deph}}/4$, 
the transport-mediated level width of $\left|\lambda_{k}\right>$ being $\Gamma_{k}=\kappa_{k}^2 \Gamma$
and the nuclear Stark shifts 
\begin{eqnarray}
\delta_{k} & = & \frac{a_{\mathrm{hf}}^{2}\epsilon_{k}}{4\left[\epsilon_{k}^{2}+\tilde{\Gamma}_{k}^{2}\right]}.
\end{eqnarray}
The non-local nuclear operators $L_{k}$ and $\mathbb{L}_{k}$ are defined as $L_{k}=\nu_{k}A_{1}^{+}+\mu_{k}A_{2}^{+}$
and $\mathbb{L}_{k}=\mu_{k}A_{1}^{-}+\nu_{k}A_{2}^{-}$, respectively. 
Finally, the last term reads 
\begin{equation}
\mathcal{K}_{\mathrm{zz}}\left[\sigma\right]=\gamma_{\mathrm{zz}}\sum_{i,j}\left[\delta A_{j}^{z}\sigma\delta A_{i}^{z}-\frac{1}{2}\left\{ \delta A_{i}^{z}\delta A_{j}^{z},\sigma\right\} \right],
\end{equation}
where $\gamma_{\mathrm{zz}}=a_{\mathrm{hf}}^2/4\gamma_{\pm}$.

\subsection{Hyperfine Coupling and Ideal Nuclear Steady State}

\begin{figure}
\includegraphics[width=1\columnwidth]{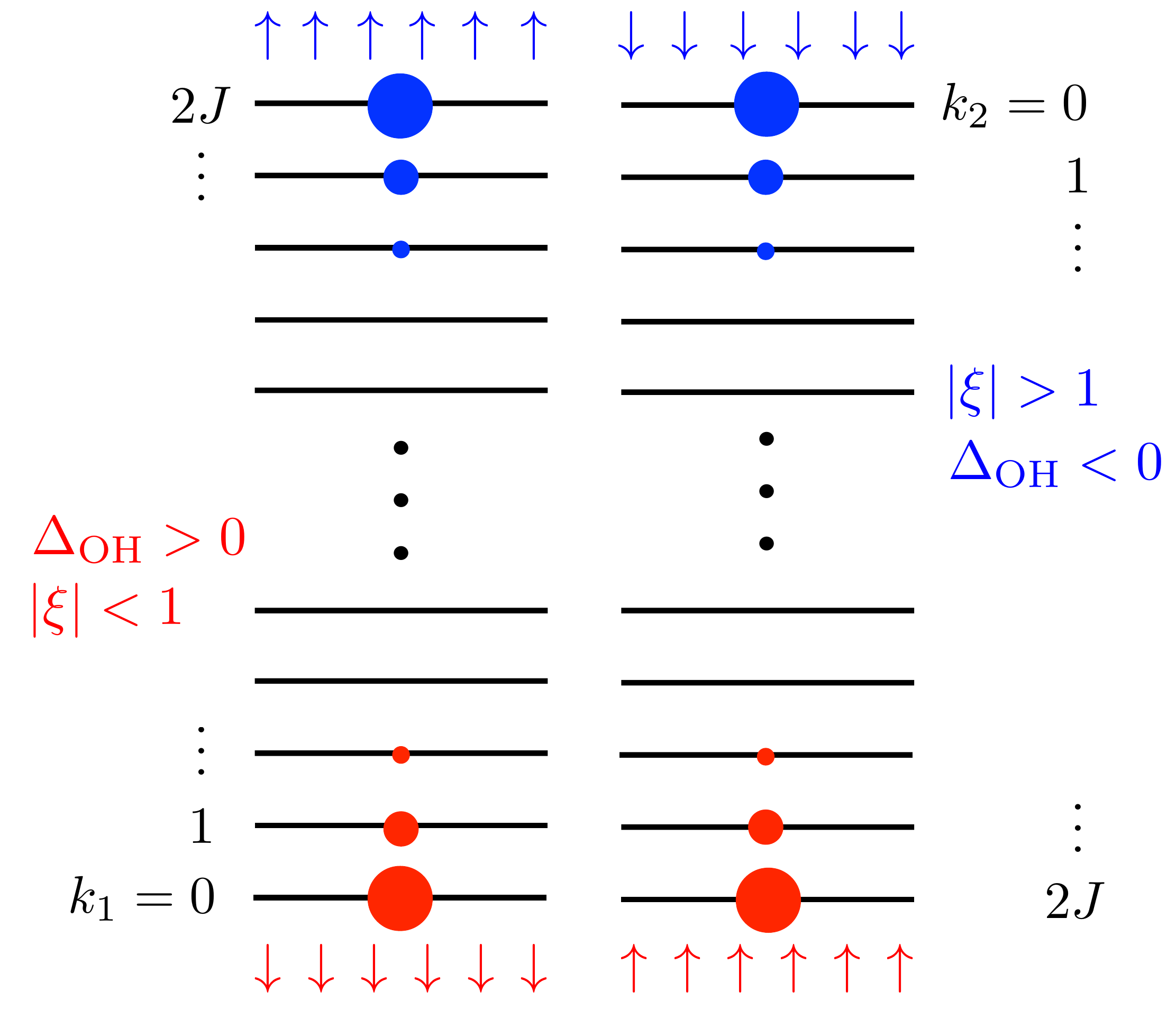}

\caption{\label{fig:dark-state-uniform-sketch}(color online). Sketch of the
ideal nuclear dark state for uniform HF coupling $\left|\xi_{\mathrm{ss}}\right\rangle $.
The Dicke states are labeled according to their spin projection $k_{i}=0,1,\dots2J$.
Since $k_{1}=k$ is strongly correlated with $k_{2}=2J-k$, the two
Dicke ladders are arranged in opposite order. The bistability inherent
to $\left|\xi_{\mathrm{ss}}\right\rangle $ is schematized as well:
The size of the spheres refers to $\left|\left<k_{1},k_{2}|\xi_{\mathrm{ss}}\right>\right|^{2}$
for $\left|\xi\right|<1$ (red) and $\left|\xi\right|>1$ (blue),
respectively. As indicated by the arrows for individual nuclear spins,
$\left|\xi\right|<1$ $\left(\left|\xi\right|>1\right)$ corresponds
to a nuclear OH gradient $\Delta_{\mathrm{OH}}>0$ $\left(\Delta_{\mathrm{OH}}>0\right)$,
respectively. }
\end{figure}

In the main text, the HP analysis has been performed 
for uniform hyperfine coupling. 
This simplification is based on the assumption that
the electron density is approximately constant in the dots and zero outside \cite{rudner11a}.
In Ref.\cite{schwager10}, 
it was shown that corrections to this idealized setting
are of the order of $1-p$ for high polarization $p$. Therefore, 
the analysis for uniform HF coupling 
is correct to zeroth order in the small parameter $1-p$.
To make connection with a realistic situation, the underlying
idea is to express the HF coupling constants as 
$a_{i,j}=\bar{a}+\delta_{i,j}$, where 
the dominant uniform term $\bar{a}$ enables an efficient description within fixed $J_{i}$ subspaces,
while the non-uniform contribution $\delta_{i,j}$ leads to a coupling 
between different $J_{i}$ subspaces on a much longer time scale. 
The latter is relevant in order to reach highly polarized nuclear states \cite{christ07}.  

We have explicitly stated the ideal nuclear steady state $\left|\xi_\mathrm{ss}\right>$,
fulfilling $L_{2}\left|\xi_\mathrm{ss}\right>=\mathbb{L}_{2}\left|\xi_\mathrm{ss}\right>=0$, 
for two 'opposing' limits:
First, we analytically construct the ideal (pure) nuclear steady-state in
the limit of identical dots $(a_{1j}=a_{2j}\forall j=1,\dots,N_{1}\equiv N_{2}=N)$
for uniform HF-coupling where $a_{ij}=N/N_{i}$. In this limit, the
non-local nuclear jump operators simplify to 
\begin{eqnarray}
L_{2} & = & \nu I_{1}^{+}+\mu I_{2}^{+},\\
\mathbb{L}_{2} & = & \mu I_{1}^{-}+\nu I_{2}^{-}.
\end{eqnarray}
Here, to simplify the notation, we have replaced $\mu_{2}$ and $\nu_{2}$
by $\mu$ and $\nu$, respectively. The common proportionality factor
is irrelevant for this analysis and therefore has been dropped. The
collective nuclear spin operators $I_{1,2}^{\alpha}$ form a spin
algebra and the so-called Dicke states 
$\left|J_{1},k_{1}\right\rangle \otimes\left|J_{2},k_{2}\right\rangle \equiv\left|J_{1},k_{1};J_{2},k_{2}\right\rangle $,
where the total spin quantum numbers $J_{i}$ are conserved and the
spin projection quantum number $k_{i}=0,1,\dots,2J_{i}$, allow for
an efficient description. Here, we restrict ourselves to the symmetric
case where $J_{1}=J_{2}=J$; analytic and numerical evidence for small
$J_{i}\approx3$ shows, that for $J_{1}\neq J_{2}$ one obtains a
mixed nuclear steady state  \cite{schuetz13}. The total spin quantum numbers $J_{i}=J$
are conserved and we set $ $$\left|J,k_{1};J,k_{2}\right\rangle =\left|k_{1},k_{2}\right\rangle $.
Using standard angular momentum relations, one obtains 
\begin{eqnarray}
L_{2}\left|k_{1},k_{2}\right\rangle  & = & \nu j_{k_{1}}\left|k_{1}+1,k_{2}\right\rangle +\mu j_{k_{2}}\left|k_{1},k_{2}+1\right\rangle ,\label{eq:L2-Dicke}\\
\mathbb{L}_{2}\left|k_{1},k_{2}\right\rangle  & = & \mu g_{k_{1}}\left|k_{1}-1,k_{2}\right\rangle +\nu g_{k_{2}}\left|k_{1},k_{2}-1\right\rangle .\label{eq:LL2-Dicke}
\end{eqnarray}
Here, we have introduced the matrix elements 
\begin{eqnarray}
j_{k} & = & \sqrt{J\left(J+1\right)-\left(k-J\right)\left(k-J+1\right)},\\
g_{k} & = & \sqrt{J\left(J+1\right)-\left(k-J\right)\left(k-J-1\right)}.
\end{eqnarray}
Note that $j_{2J}=0$ and $g_{0}=0$. Moreover, the matrix elements
obey the symmetry 
\begin{eqnarray}
j_{k} & = & j_{2J-k-1},\\
g_{k+1} & = & g_{2J-k}.
\end{eqnarray}
Now, we show that $\left|\xi_{\mathrm{ss}}\right\rangle $ fulfills
$L_{2}\left|\xi_{\mathrm{ss}}\right\rangle =\mathbb{L}_{2}\left|\xi_{\mathrm{ss}}\right\rangle =0$.
First, using the relations above, we have 
\begin{eqnarray*}
L_{2}\left|\xi_{\mathrm{ss}}\right\rangle  & = & \sum_{k=0}^{2J}\xi^{k}\left[\nu j_{k}\left|k+1,2J-k\right\rangle \right.\\
 &  & \left.+\mu j_{2J-k}\left|k,2J-k+1\right\rangle \right]\\
 & = & \sum_{k=0}^{2J-1}\xi^{k}\left[\nu j_{k}\left|k+1,2J-k\right\rangle \right.\\
 &  & \left.+\xi\mu j_{2J-k-1}\left|k+1,2J-k\right\rangle \right]\\
 & = & \sum_{k=0}^{2J-1}\xi^{k}\nu\underset{=0}{\underbrace{\left[j_{k}-j_{2J-k-1}\right]}}\left|k+1,2J-k\right\rangle .
\end{eqnarray*}
In the second step, since $j_{2J}=0$, we have redefined the summation
index as $k\rightarrow k+1$. Along the same lines, one obtains 
\begin{eqnarray*}
\mathbb{L}_{2}\left|\xi_{\mathrm{ss}}\right\rangle  & = & \sum_{k=0}^{2J}\xi^{k}\left[\mu g_{k}\left|k-1,2J-k\right\rangle \right.\\
 &  & \left.+\nu g_{2J-k}\left|k,2J-k-1\right\rangle \right]\\
 & = & \sum_{k=0}^{2J-1}\xi^{k}\left[\xi\mu g_{k+1}\left|k,2J-k-1\right\rangle \right.\\
 &  & \left.+\nu g_{2J-k}\left|k,2J-k-1\right\rangle \right]\\
 & = & \sum_{k=0}^{2J-1}\xi^{k}\nu\underset{=0}{\underbrace{\left[g_{2J-k}-g_{k+1}\right]}}\left|k,2J-k-1\right\rangle .
\end{eqnarray*}
This completes the proof. For illustration, the dark state $\left|\xi_{\mathrm{ss}}\right\rangle $
is sketched in Fig. \ref{fig:dark-state-uniform-sketch}. In particular,
the bistable polarization character inherent to $\left|\xi_{\mathrm{ss}}\right\rangle $
is emphasized, as (in contrast to the bosonic case) the modulus of
the parameter $\xi$ is not confined to $\left|\xi\right|<1$. 

\begin{figure}
\includegraphics[width=1\columnwidth]{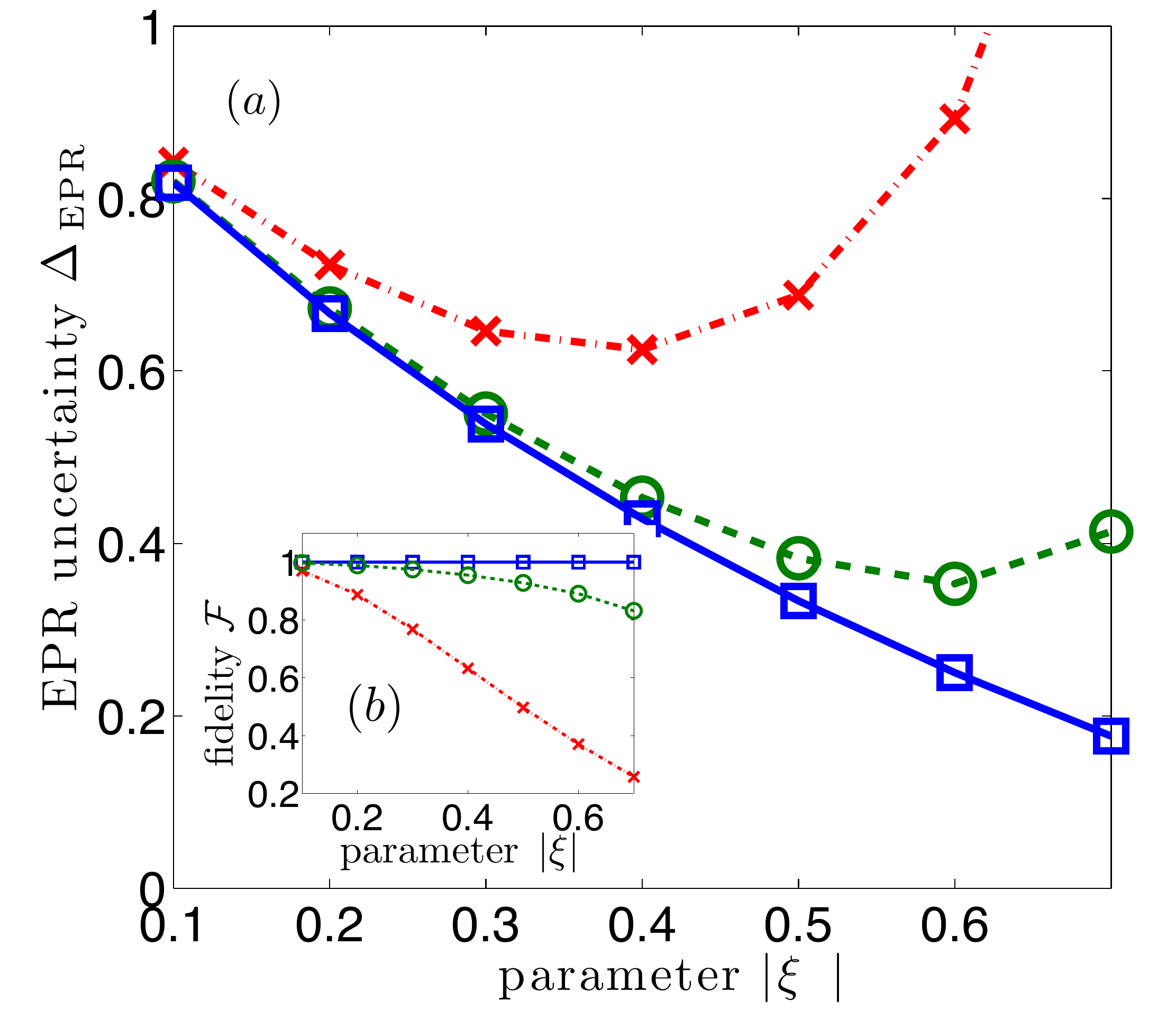}

\caption{\label{fig:EPR-fidelity-inhomog}(color online). EPR uncertainty $\Delta_{\mathrm{EPR}}$
(a) and fidelity $\mathcal{F}$ with the nuclear target state $\left|\xi_{\mathrm{ss}}\right>$ (b) as a 
function of the squeezing parameter $\left|\xi\right|$ for $N_{1}=N_{2}=3$ inhomogeneously coupled
nuclear spins. The blue curve (squares) refers to an symmetric setting where 
$\vec{a}_{1}=\vec{a}_{2}=\left(1.11,1.67,0.22\right)$, whereas the green (circles) and red (crosses)
curves incorporate asymmetries: $\vec{a}_{1}=\left(1.18,1.61,0.21\right)$, $\vec{a}_{2}=\left(1.11,1.67,0.22\right)$
and $\vec{a}_{1}=\left(1,1.5,0.5\right)$, $\vec{a}_{2}=\left(1.24,1.55,0.21\right)$, respectively.}
\end{figure}

Second, we have elaborated on the case 
of a perfectly inhomogeneous distribution of HF coupling constants. 
For identical QDs, the nuclear spins can always be grouped into pairs
$\left(a_{1,j}=a_{2,j}\right)$. 
In the absence of  degeneracies, i.e.,
for $a_{i,j} \neq a_{i,k}$ for all $j \neq k$, we have identified the nuclear dark state as
$\left|\xi_{\mathrm{ss}} \right\rangle =\otimes_{j=1}^{N}\left|\xi\right\rangle_{j} $. 
This analytical result is verified by exact diagonalization for small systems
of inhomogeneously coupled nuclear spins $\left(N_{1}=N_{2}=3\right)$: see  \reffig{fig:EPR-fidelity-inhomog}. 
It indicates that $\left|\xi_{\mathrm{ss}} \right\rangle$ is 
the unique steady state. 
Moreover, as long as the squeezing parameter is $\left|\xi\right| \lesssim 0.5$,
the nuclear system is found to be robust against 
asymmetries $\left(\vec{a}_{1} \neq \vec{a}_{2}\right)$
and features entanglement over a broad range of the parameter $\left|\xi\right|$.  

Note that one can 'continuously' go from the case of non-degenerate HF
coupling constants to the limit of uniform HF coupling by 
grouping spins with the same HF coupling constants to 'shells', 
which form collective nuclear spins.    
For degenerate couplings, however, there are additional conserved
quantities and therefore multiple stationary states of the above form. 
If $a_{1,j}\approx a_{2,j}$ we expect (and have also verified for small $N$, see  \reffig{fig:EPR-fidelity-inhomog})
that the resulting mixed stationary state is still unique (in the
non-degenerate case) and close to $\Ket{\xi_\mathrm{ss}}$.

\subsection{Holstein-Primakoff Transformation}
The (exact) Holstein-Primakoff transformation expresses the truncation of the collective nuclear
spin operators to a total spin $J_{i}$ subspace in terms of a bosonic mode \cite{kessler12}. 
For $\Delta>0 \left(\left|\xi\right|<1\right)$ the nuclear ensembles are polarized in opposite directions, and 
the (zeroth order) HP mapping for the collective nuclear spins $I_{i}^{\alpha}=\sum_{j}\sigma_{i,j}^{\alpha}$ 
$(\alpha=\pm,z)$
reads explicitly 
\begin{eqnarray}
I_{1}^{-} & \approx & \sqrt{2J_{1}}b_{1},\\
I_{1}^{z} & = & b_{1}^{\dagger}b_{1}-J_{1}.
\end{eqnarray}
for the first nuclear ensemble, and similarly for the second ensemble
\begin{eqnarray}
I_{2}^{+} & \approx & \sqrt{2J_{2}}b_{2},\\
I_{1}^{z} & = & J_{2}-b_{2}^{\dagger}b_{2}.
\end{eqnarray} 
We consider the subspace with large collective spin quantum numbers, that is
$J_{i}\sim\mathcal{O}\left(N/2\right)$. Thus, the zeroth-order HP mapping given above can be justified
self-consistently, provided that the occupations in the bosonic modes $b_{i}$
are small compared to $2J_{i}$ \cite{kessler12}.

For equal dot sizes and $J_{1}=J_{2}=J$, the nuclear jump
operators are mapped to $L_{2}\sim a$ and $\mathbb{L}_{2}\sim \tilde{a}$,
where $a=\mu b_{2}+\nu b_{1}^{\dagger}$ and $\tilde{a}=\mu b_{1}+\nu b_{2}^{\dagger}$.
Here, we have set $\mu=\mu_{2}/\sqrt{\mu_{2}^2 - \nu_{2}^2}$ and similarly for $\nu$ 
such that $\mu^{2}-\nu^{2}=1$.
In this picture,  
the (unique) ideal steady state is well-known to be a two-mode squeezed
state 
\begin{eqnarray}
\left|\Psi_{\mathrm{TMS}}\right\rangle =\mu^{-1}\sum_{n}\xi^{n}\left|n,n\right\rangle,
\end{eqnarray}
which is simply the vacuum in the \textit{non-local} bosonic modes $a$ and
$\tilde{a}$ fulfilling $a\left|\Psi_{\mathrm{TMS}}\right\rangle =\tilde{a}\left|\Psi_{\mathrm{TMS}}\right\rangle =0$ \cite{muschik11}. 

The generation of entanglement can be certified via 
the EPR entanglement condition \cite{muschik11,raymer03},
where the EPR-uncertainty is given by
\begin{eqnarray}
\Delta_{\mathrm{EPR}} & = & \Sigma_{J}/\left(\left|\left<I_{1}^{z}\right>\right|+\left|\left<I_{2}^{z}\right>\right|\right) \\
 & = & \frac{1}{2} \left[\mathrm{var}\left(X_{1}+X_{2}\right)+\mathrm{var}\left(P_{1}-P_{2}\right)\right],
\end{eqnarray} 
where 
$\Sigma_{J}=\mathrm{var}\left(I_{1}^{x}+I_{2}^{x}\right)+\mathrm{var}\left(I_{1}^{y}+I_{2}^{y}\right)$.
Here, 
\begin{eqnarray}
X_{i} & = & (b_{i}+b_{i}^{\dagger})/\sqrt{2},\\
P_{i} & = & i(b_{i}^{\dagger}-b_{i})/\sqrt{2}.
\end{eqnarray} 
refer to the quadrature operators related to the local bosonic modes $b_{i}$.

\subsection{EPR Uncertainty}

Within the HP approximation, the evaluation of $\Delta_{\mathrm{EPR}}$ 
is based on the standard form of the
steady state covariance matrix, defined as
$\Gamma^{\mathrm{CM}}_{ij}=\left\langle \left\{ R_{i},R_{j}\right\} \right\rangle -2\left\langle R_{i}\right\rangle \left\langle R_{j}\right\rangle $,
where $\left\{ R_{i},i=1,\dots,4\right\} =\left\{ X_{1},P_{1},X_{2},P_{2}\right\} $.
Up to \textit{local} unitary operations, $\Gamma^{\mathrm{CM}}$ can always 
be written in standard form
\begin{equation}
\Gamma^{\mathrm{std}}=S^{\top}\Gamma^{\mathrm{CM}}S=\left(\begin{array}{cccc}
a & 0 & k_{1} & 0\\
0 & a & 0 & k_{2}\\
k_{1} & 0 & b & 0\\
0 & k_{2} & 0 & b
\end{array}\right).
\end{equation}

\textit{Squeezing parameter.---}The amount of entanglement can be tuned via 
the squeezing parameter $\xi$. For fixed $\epsilon>0$, $\Delta>0$ and increasing
tunneling parameter $t$, 
$\epsilon_{2}$ approaches 0 [compare Fig.1 (b) in the main text], so that 
the relative weight of $\nu_{2}$ 
as compared to $\mu_{2}$ increases. 
This results in a larger squeezing parameter $|\xi|=|\nu_{2}/\mu_{2}|$.

\subsection{Polarization Dynamics}

Starting out from Eq.(\ref{eq:EOM-nuclear-spins-1}) we obtain 
dynamical equations for the nuclear polarizations $\left<I_{i}^{z}\right>$.  
For simplicity, we then employ a semiclassical factorization scheme, which neglects correlations among 
different nuclear spins by setting $\left<\sigma_{i}^{+}\sigma_{j}^{-}\right>=\left<\sigma_{i}^{z}\right>+1/2$ 
for $i=j$ and zero otherwise
(note that $\Ket{\xi_{\mathrm{ss}}}$ tends to a maximally polarized product state for $\left|\xi\right| \rightarrow 0$). 
This zeroth-order approximation directly leads to a closed equation of motion 
for $\Delta_{I^{z}}$ as stated in Eq.(5) in the main text.
Here, we have introduced the effective HF-mediated pumping rate $\chi$ 
and depolarization rate $\gamma_{\mathrm{eff}}$ as 
\begin{eqnarray}
\chi & = & \gamma\left(\mu_{2}^{2}-\nu_{2}^{2}\right)\left(3p_{+}-1\right),\label{eq:semiclassical-depolarization-rate-gamma-eff}\\
\gamma_{\mathrm{eff}}  & = & \gamma\left(\mu_{2}^{2}+\nu_{2}^{2}\right)\left(1-p_{+}\right).\label{eq:semiclassical-pumping-rate-chi}
\end{eqnarray}
Note that according to Eq.(5) the nuclear fixed point 
polarization gradient
is proportional to the ratio 
$\left(\mu_{2}^{2}-\nu_{2}^{2}\right) / \left(\mu_{2}^{2}+\nu_{2}^{2}\right) = \left(1- \xi^2 \right)/ \left(1+ \xi^2 \right)$.
This coincides with the nuclear polarization gradient inherent to the dark state $\left|\xi_{\mathrm{ss}}\right>$.
Accordingly, Eq.(5) in the main text can be reformulated as 
\begin{eqnarray}
\frac{d}{dt}\Delta_{I^{z}} & = & -\gamma_{\mathrm{eff}}\left[\Delta_{I^{z}}-N\frac{1-\xi^2}{1+\xi^2}\frac{3p_{+}-1}{1-p_{+}}\right].\label{eq:EOM-Delta-Iz-semiclassical-closed-2}
\end{eqnarray}
Here, the last factor is one in the high gradient regime where $p_{+}=1/2$,  
but may suppress high polarization solutions in the low gradient 
regime $\left(p_{+} \approx 1/3\right)$ \cite{schuetz13}.

\textit{Time scales}.---As shown in Fig.3, we can estimate $\dot{\Delta}_{I^{z}} \approx 0.1\mathrm{MHz}$. 
In order to reach a highly polarized fixed point, approximately $\sim 10^{5}$ nuclear spin flips
are required; therefore, the total time for the polarization process is approximately
$\sim 10^5/ 0.1\mathrm{MHz} \approx 1\mathrm{s}$. 
This is in agreement with typical time scales observed in nuclear polarization experiments \cite{takahashi11}.
Lastly, $\gamma_{\mathrm{eff}}^{-1} \approx 1 \mathrm{s}$ is compatible with the semiclassical
approximation, since nuclear spins typically dephase at a rate of $\sim \mathrm{kHz}$ \cite{gullans10, takahashi11}. 

\subsection{External Magnetic Fields}

In the main text we have assumed $\omega_{\mathrm{ext}}=\Delta_{\mathrm{ext}}=0$ for simplicity. 
As shown here, non-vanishing external fields do not lead to qualitative
changes in the in the principal effects.
First, the presence of a non-vanishing external gradient $\Delta_{\mathrm{ext}}$ is actually beneficial 
for our scheme as it can provide an efficient way in order to kick-start the nuclear self-polarization process. 
Second, a non-zero homogeneous external field $\omega_{\mathrm{ext}}$ leads to 
a non-zero splitting between the Pauli-blocked triplets $\left|T_{\pm}\right>$. 
This gives rise to an asymmetry in the effective HF-mediated quantities $\gamma$ and $\delta$, as the 
detunings for the transitions from $\left|T_{\pm}\right\rangle$ to $\left|\lambda_{2}\right\rangle$ (and vice versa) 
are different for $\omega_{0}\neq0$. Importantly, however, the kernel of $\mathcal{L}_{\mathrm{id}}$, 
associated with the ideal steady state, is unaffected by this asymmetry. 

\subsection{Summary of Experimental Requirements}

Here, we summarize the requirements for an experimental realization of our scheme:
The condition $t \gg \omega_{0},g_{\mathrm{hf}}$ ensures that the Pauli blockade 
is primarily lifted via the electronic level $\left|\lambda_{2}\right>$. Then, $\Delta\gtrsim 3 \mu\mathrm{eV}$, 
together with $\Gamma \gg \gamma_{\pm},\gamma_{\mathrm{deph}} \gg g_{\mathrm{hf}}$,
guarantees that the electronic system settles into the desired quasi steady state
on a time scale much shorter than the nuclear dynamics. 
As shown in Sec.(\ref{the-model}), the latter could be realized by, e.g.,  working in a regime of efficient cotunneling processes.
To kick-start the nuclear self-polarization process towards a high-gradient stable fixed point, 
some initial gradient of approximately $\sim (1-2) \mu\mathrm{eV}$ is required.  
Finally, in order to beat nuclear spin decoherence, one needs 
$t^{*} \ll 1\mathrm{ms}$.

All these requirements can be met simultaneously in a quantum dot 
defined in a two-dimensional
GaAs/AlGaAs electron gas by
a pattern of Schottky gates fabricated on the surface 
with electron beam lithography; see e.g. Ref.\cite{hanson07}. This approach for realizing quantum dots 
has proven to be extremely powerful, since many of the relevant parameters can be tuned in-situ.

Due to the exponential dependence of tunnel coupling strength on gate voltage, all the tunnel 
barriers can be varied
from less than $10^{-12} \mathrm{eV}$ (a millisecond timescale, verified 
by real-time detection of single charges hopping on or off the dot) to about $100 \mu \mathrm{eV}$ 
(verified by the broadening of the time-averaged charge transition; 
note that for much larger tunnel couplings, two 
neighbouring dots become one single dot). This extreme tunability applies to the interdot barrier characterized by $t$, 
as well as to the dot-lead barriers characterized by $\Gamma$.

The detuning $\epsilon$ between the dots can be varied anywhere between zero and a positive or 
negative detuning equal to the addition energy, at which point additional electrons are pulled into the dot. 
The typical energy scale for the addition energy is $1-3 \mathrm{meV}$.

Less choice exists in the parameters related to the electron-nuclear spin interaction, so in the analysis 
we used the typical numbers \cite{hanson07}. In particular, in typical dots, the electron is in contact with $N \sim 10^{6}$ nuclei. 
Also fixed is the total electron-nuclear coupling strength $A_{\mathrm{HF}} \sim 100 \mu\mathrm{eV}$. 
$N$ and  $A_{\mathrm{HF}} $ together set $g_{\mathrm{hf}} \sim 0.1 \mu\mathrm{eV}$. 
Finally, the nuclear spin coherence time of $\sim 1 \mathrm{ms}$ is fixed as well \cite{takahashi11}.

The extreme tunability of the electronic parameters $t$ and (in particular) $\Gamma$ allows us 
to reach the desired regime, where the electronic system quickly settles into its mixed quasi steady state on 
relevant nuclear time scales. As shown in more detail in Sec.(\ref{the-model}),
one can make the dissipative mixing and dephasing rates (which are both proportional to $\Gamma$) 
fast compared to $g_{\mathrm{hf}}$ by going to a regime of efficient electron exchange with the reservoirs.


\begin{thebibliography}{10}
\bibitem{Gamma} For fast recharging of the DQD, 
$\Gamma=\Gamma_{R}/2$, where $\Gamma_{R}$ is the 
sequential tunneling rate to the right lead \cite{SuppInfo}.
\end{thebibliography}
\end{document}